%% file: mutual-contact-discovery.tex
\RequirePackage[l2tabu, orthodox]{nag}
\documentclass[a4paper,11pt]{article}
\input mutual-contact-discovery-includes.tex
\input mutual-contact-discovery-macros.tex
\title{Mutual Contact Discovery\thanks{%
    Previous versions of this paper overstate the properties of the protocol and contain an erroneous proof.\\
    Version: Tue Dec 5 13:35:13 2023 +0100 / arxiv-v4 / mutual-contact-discovery.tex    
}}
\author{Jaap-Henk Hoepman\\
  Radboud University Nijmegen\\
  Karlstad University \\
  University of Groningen\\
  (\email{jhh@cs.ru.nl})
}
\begin{document}
\maketitle
\input mutual-contact-discovery-abstract.tex
\input mutual-contact-discovery-body.tex

%
%
%
\setlength{\emergencystretch}{8em}
\printbibliography
\appendix
\input mutual-contact-discovery-appendix.tex
\end{document}

%% file: mutual-contact-discovery-includes.tex
\let\email=\relax
\usepackage[charter]{mathdesign}

\usepackage[T1]{fontenc}
\usepackage[utf8]{inputenc}
\usepackage{textcomp} 
\PassOptionsToPackage{hyphens}{url}
\usepackage[bookmarksnumbered,breaklinks,colorlinks=true]{hyperref}
\usepackage[hyphenbreaks]{breakurl}
\usepackage{graphicx}
\usepackage{amsmath}
\usepackage[english]{babel}
\usepackage[style=numeric,maxnames=5,giveninits=true,doi=false,isbn=false,url=true,eventdate=comp]{biblatex}
\usepackage[short]{bibstrings}
\usepackage{xspace}
\usepackage{verbatim}
\usepackage{general}
\usepackage{markup}
\usepackage{environment}
\usepackage{hyphenation}

\AtEveryBibitem{%
  \ifentrytype{inproceedings}{
    \clearfield{url}
    \clearname{editor}%
    \clearfield{booktitleaddon}%
  }{}
  \ifentrytype{article}{
    \clearfield{url}
  }{}
}

\DeclareFieldFormat[misc]{title}{\mkbibquote{#1}}
\DeclareSourcemap{
  \maps{
    \map{
      \step[typesource=tweet, typetarget=misc, final]
      \step[fieldset=howpublished, fieldvalue={Twitter}]
    }
    \map{
      \step[typesource=blog, typetarget=misc, final]
      \step[fieldset=entrysubtype, fieldvalue={blog}]
      }
    \map{
      \step[typesource=movie, typetarget=misc, final]
      }
    \map{
      \step[typesource=newspaper, typetarget=article, final]
      \step[fieldset=entrysubtype, fieldvalue={newspaper}]
      }
    }
  }

%% file: mutual-contact-discovery-macros.tex
\newcommand{\secpar}{\sigma}

\newcommand{\csym}{\lhd}
\newcommand{\cset}{S_{\csym}}
\newcommand{\contacts}[1]{\mathit{contacts}_{#1}}
\newcommand{\hidden}[1]{\mathit{hiddenby}_{#1}}
\newcommand{\visible}[1]{\mathit{visible}_{#1}}

\newcommand{\mutualcontact}[2]{(#1 \in \visible{#2}) \wedge (#2 \in \contacts{#1})}

\newcommand{\mcsym}{\rtimes}

\newcommand{\mcset}{S_{\mcsym}}

\newcommand{\mc}[2]{#1 \mcsym #2}

\newcommand{\proofpart}[1]{\medskip\noindent\emph{#1}:\par\medskip\noindent}

\newcommand{\sgu}{V}
\newcommand{\sge}{E}

\newcommand{\members}{\mathcal{M}}
\newcommand{\matchingserver}{\mathcal{X}}
\newcommand{\trustedserver}{\mathcal{T}}

\newcommand{\out}[1]{\mathit{out}_{#1}}

\newcommand{\mcdsetup}{\mathsf{Setup}}
\newcommand{\mcddiscover}{\mathsf{Discover}}

\newcommand{\adversary}[1]{\mathcal{A}_{\mathrm{#1}}}

\newcommand{\simulator}[1]{\mathcal{S}_{#1}}
\newcommand{\prot}{\Pi}
\newcommand{\ideal}{I}
\newcommand{\err}{\epsilon}

\newcommand{\G}[1]{\mathbb{G}_{#1}}
\newcommand{\pairingsym}{\mathbf{e}}
\newcommand{\pairing}[2]{\pairingsym(#1,#2)}
\renewcommand{\H}[1]{H_{#1}}
\newcommand{\HT}{\H{2}^{<}}
\newcommand{\point}[1]{P_{#1}}
\newcommand{\cert}[1]{C(#1)}
\newcommand{\cred}[1]{Z(#1)}
\newcommand{\token}[2]{T_{#1#2}}
\newcommand{\Ppub}{P_{\mathrm{pub}}}
\newcommand{\tokenx}[4]{\HT(\token{#1}{#2},\point{#3},\point{#4})}

\newcommand{\concat}{\,\|\,}
\newcommand{\sym}[1]{\mathsf{#1}}
\newcommand{\KDF}[1]{\sym{KDF}(#1)}
\newcommand{\tuple}[1]{\langle #1 \rangle}
\newcommand{\mutuall}[2]{v_{#1#2}}
\newcommand{\mutualwork}[2]{\H{2}(\token{#1}{#2},P)}

\newcommand{\mutualworkB}[2]{\H{2}(\token{#1}{#2},#2)}

%% file: mutual-contact-discovery-abstract.tex
\begin{abstract}{
  Contact discovery allows new users of a messaging service to find existing contacts that already use that service. Existing users are similarly informed of new users that join. 
  This creates a privacy issue: anyone already on the service that has your number on their contact list gets notified that you joined. Even if you don't know that person, or if it is an ex or former colleague that you long parted with and whose contact details you deleted long ago.
  To solve this, we propose a \emph{mutual} contact discovery protocol, that only allow users to discover each other when \emph{both} are (still) in each other's contact list. Mutual contact discovery has the additional advantage that it can be implemented in a more privacy friendly fashion (\eg protecting the social graph from the server) than traditional, one-sided contact discovery, without necessarily relying on trusted hardware.
}
\end{abstract}

%% file: mutual-contact-discovery-body.tex
\section{Introduction}
\label{sec-intro}

Many messaging services like WhatsApp\footnote{%
  \url{https://www.whatsapp.com}
}
or Signal\footnote{%
  \url{https://signal.org}
}
are based on the observation that ``access to an existing social graph makes building social apps much easier''~\autocite{moxie2014contact-discovery}, and thus use an existing set of globally unique identifiers (typically mobile phone numbers) to identify their users. This allows such services to notify their users when any of their existing contacts join the service. To implement this \term{contact discovery} process, users need to allow the messaging service to access the contact list or address book they maintain locally on their smartphone (or other device), so that the service can match any entry on this list with the list of new users that recently joined. Contact discovery creates privacy issues: a naive implementation (where a user asks the service provider whether any of the phone numbers on their contact list happen to be subscribed as well) allows the service provider to learn the mobile phone number of all contacts of all its users, even when these contacts are not member of the service at all~\autocite{hagen2021contact-discovery}.
This is in particular a concern for people that shield their mobile phone number for privacy reasons.
As a result, messaging services like Signal try to make the discovery process more privacy friendly by limiting the amount of information they learn~\autocite{moxie2017contact-discovery}. As we will discuss in section~\ref{sec-sota}, the protocols currently in use are somewhat limited in the protection they offer, are not very efficient in practice, or rely on non-standard assumptions.

When users switch to a supposedly privacy friendly messaging app like Signal, they are surprised to find out that contact discovery is \emph{one-sided}. When you enable contact discovery, anyone already on Signal that has your number on their contact list (like a former colleague, former patient, or an ex that you long parted with but has kept your phone number as a contact) gets notified that you joined.\footnote{%
  See \eg 
  \url{https://twitter.com/DorotheaBaur/status/1349340273357291528} .
}
Privacy conscious people expect a somewhat different behaviour: people should only be allowed to discover that someone else is on a messaging service when \emph{both of them} have the phone number of the other person on their contact list. We call this \term{mutual contact discovery}. In mutual contact discovery, the fact that you have someone else's number in your contact list is only a signal that you allow this person to discover you. Knowing someone's number and adding it to your contact list does \emph{not} give you immediate permission to discover the other person: this only happens if that person also has \emph{your} number on \emph{their} contact list (and when both of you have contact discovery enabled, of course).

Note that more fine-grained approaches for mutual contact discovery are very important in practice. For example, users that have a phone number in their contact list to recognise a harasser calling, may want to have the option to disallow such numbers from discovering them. Conversely, people deleting contacts from their contact list may not want the deleted contact to be actively notified about this. We leave these refinements for further study, but note that such options can easily be added to the protocols discussed in this paper.

In this paper we study privacy friendly protocols for mutual contact discovery. We first discuss existing schemes for (one-sided) contact discovery and related problems like private matchmaking and private set intersection in section~\ref{sec-sota}, and show how these do not satisfactorily solve the problem at hand. We then set the stage in section~\ref{sec-model}, describing the system model, threat model, security and privacy requirements, and primitives used. After that we present our main protocol for mutual contact discovery in section~\ref{sec-protocol}. This protocol offers stronger privacy guarantees in a more adversarial model than typically assumed: only mutual contacts can discover each other, and the matching server cannot learn the identities of non-members, cannot reconstruct part of the social graph and does not even learn the size of the contact lists of individual users. The protocol is not yet perfect however: a malicious mutual contact may prevent itself from being discovered. The protocol assumes sender anonymous communication channels, and uses the existing social graph as the source for certified identifiers. This is a stronger assumption than typically supposed by existing contact discovery protocols, but we will argue  this is still a reasonable assumption to make in section~\ref{ssec-certs}. We discuss our results and summarise our conclusions in section~\ref{sec-concl}.

\section{State of the art}
\label{sec-sota}

\subsection{Traditional, one-sided, contact discovery}

Many messaging services implement the traditional one-sided version of contact discovery where you get immediately notified when anyone on your contact list is a member as well.

WhatsApp uses a particularly naive and privacy invasive protocol. Its optional contact upload feature uploads phone numbers in plaintext to the WhatsApp server to identify contacts that are members as well. It stores hashes of those phone numbers that are not a member yet to ``to more efficiently connect you with these contacts if they join WhatsApp''.\footnote{%
  \url{https://faq.whatsapp.com/general/contacts/about-contact-upload}, accessed 2021-10-14.
}

A natural candidate to make one-sided contact discovery more privacy friend-ly is to use private set intersection~\autocite{freedman2004setintersect} to compute the intersection of the contact list (held by the user) and the total set of users of the messaging service (held by the server). These academic approaches are not practical however, as the traditional protocols do not take the unbalanced nature of the sets involved in contact discovery into account: the user has at most several hundreds of contacts, whereas the service typically has billions of users~\autocite{moxie2014contact-discovery}. Even private set intersection (PSI) protocols optimised for unbalanced sets~\autocite{kiss2017psi-for-contact-discovery} perform poorly even after extensive optimisation: matching a thousand contacts against a quarter billion registered users still takes three seconds (assuming a WiFi connection)~\autocite{kales2019private-contact-discovery-at-scale}. Such protocols may optimise the computational load for the device with the smallest set, but typically at the expense of requiring complex operations to be performed by the server storing the larger set. Moreover, such a PSI protocol needs to be run with \emph{each} of the billions of users, and \eg all of the protocols discussed by Kiss \etal~\autocite{kiss2017psi-for-contact-discovery} require the server to send a significant amount of data (in the best case a Bloom filter with a constant size, chosen between $1.8$-$5.5$ MB, representing the list of known users). Other efficient PSI protocols like~\autocite{chen2017psi,chen2018psi} assume a stronger honesty requirement than usual: they allow an adversarial sender (the server) to force the output of the receiver (the mobile device of the user) to equal all the members of its set \emph{without needing to guess} the members of the set held by the mobile device as is typically the case in the honest-but-curious setting for private set intersection. This makes such protocols not useful for our application, as we wish to develop protocols that withstand even such malicious behaviour.

As result, privacy friendly messaging services like Signal resort to using trusted hardware based approaches. The basic protocol (hash the entries in a user's contact list before sending it to the server) is kept verifiably honest (in the sense that it doesn't store any information after running the contact discovery protocol) by running the server code on trusted hardware~\cite{moxie2017contact-discovery}. Clearly this is the most efficient way (at least from the client perspective) to solve contact discovery. It does require the server to have trusted hardware, and crucially depends on this hardware to be indeed trusted for privacy to hold. 

\subsection{Recent advances in mutual contact discovery}

Chaum~\etal\autocite{chaum2022udm} recently and independently\footnote{%
  The first version of this paper appeared in September 2022.
}
published a description and analysis of what they call a User Discovery with Minimal information disclosure (UDM) protocol. This protocol, invented by Chaum in 2016 but not published back then, is in essence a mutual contact discovery protocol with less strong security properties than the protocol described in this paper. It also lacks a formal security analysis. It has strong similarities with our protocol using a public key server (described in the appendix in section~\ref{sec-keyserver}), with the same limitation: requests to the key server reveal an interest in establishing contact. Also, the UDM protocol should use an Authenticated Encryption scheme in order to prevent the matching server (called the Encrypted ID Manager in the UDM protocol) from convincing users of mutual contacts that in fact do not exist. (Note that this is not a requirement in the original UDM problem statement.) An additional feature of the UDM protocol is that users can securely exchange private contact identifiers through which they can be reached on the new platform. 

Also independently of this work, Mohnblatt~\etal\autocite{mohnblatt2023arke} improve on the UDM protocol in two ways. Their Arke protocol distributes the key server (that provides users \emph{once} with their own private key, similar to how our protocol presented below uses certified identifiers; it thus avoids leaking interest in establishing contact), and it implements the matching server as a public bulletin board that is maintained in a distributed fashion using a distributed ledger (somewhat similar to our earlier work on anonymous messaging using a public bulletin board~\autocite{hoepman2015bulletinboard}). They also provide security proofs. They also claim that the performance of Arke is independent of the total number of users, but this is debatable (and also not substantiated in the paper itself). The size of the bulletin board depends on the number of participating users in each epoch. As many users may have undiscovered contacts in their contact list after an epoch, this means they will participate in every epoch.

Our protocol improves UDM~\autocite{chaum2022udm} and provides a security proof. Our protocol makes use of the fact that messaging services have a central component anyway, and thus turns out to be simpler than Arke~\autocite{cryptoeprint:2023/1218}. It uses similar, but independently discovered, core ideas though. As a result, the performance is comparable, with a similar dependence (or independence, depending how you want to frame this) on the total number of users.

\subsection{Other related protocols}

A problem very similar to mutual contact discovery, called \emph{private matchmaking}, was studied almost forty years ago by Baldwin and Gramlich \autocite{baldwin1985matchmaking}. In abstract terms, the problem there is to determine whether two \emph{known or identified} users share the same secret value (which potentially is from a small set, and therefore could be easy to guess). Potential use cases are to determine whether two agents have the same level of clearance, or a high-level job referral service where an intermediary allows a potential candidate for a particular job to verify whether a particular company is actually offering it. Baldwin and Gramlich propose a protocol involving the help of a central matchmaking server to allow two known parties $A$ and $B$ to verify whether they share the same secret. Due to certain technical details their protocol cannot be turned into a mutual contact discovery protocol directly.\footnote{%
  In essence the problem is that as it stands the Baldwin and Gramlich protocol assumes the parties can exchange messages directly, among each other, before being discovered. In other words, it assumes that the underlying source of identifiers used to establish contact is itself a kind of communication network that allows the exchange of messages. When using mobile phone numbers as identifiers this is indeed the case, but we want our protocols to be more generally applicable using different kinds of identifiers without making such a strong assumption.
}
But their ideas have inspired the protocols discussed in this paper.

Mutual contact discovery is also related to the following problems studied in the past. There is an important distinction, however, as mutual contact discovery assumes the participants are far apart and thus involves the cooperation of central matching server, while the protocols discussed below allow participants that are in each other's vicinity to discover or authenticate each other directly.

\emph{Private handshaking} protocols (that could be considered a variant of private matchmaking protocols where the secret to check is known to have large entropy and therefore cannot be guessed) are an example~\autocite{meadows1986matchmaking,balfanz2003handshake,hoepman2007privatehandshakes}.
Because these protocols assume an input from a large entropy domain, these cannot directly be used to solve mutual contact discovery (where the input, for example phone numbers,
has relatively low entropy). Moreover, these protocols assume that the parties wishing to ``shake hands'' do so directly, without the involvement of a third party. We do note that the underlying techniques used in protocols for private matching and private handshaking prove to be useful to implement mutual contact discovery as well.


Heinrich \etal\autocite{heinrich2021airdrop} studied mutual contact discovery in the context of AirDrop, Apple's peer-to-peer file exchange protocol. In AirDrop, users can choose to be only discoverable for known contacts. Apple's protocol has some privacy weaknesses, that allow an attacker to recover the contact details of both senders and receivers. Heinreich \etal propose to fix this by essentially running two private set intersection protocols with the set of known contacts and the identity (or identities) of the other participant as input.\footnote{%
  We note that a different approach requiring only one run of a private set intersection protocol (see \eg~\autocite{freedman2004setintersect}) is also possible. Assume identities can be globally ordered. Concatenate the entries within your contact list with your own identity, putting the one that comes first in the global identity order first. This way, if Alice has Bob in her contact list and vice versa, both will construct an entry ``Alice $\|$ Bob'' in the augmented contact list. Private set intersection will find a match if and only if both are member of the other person's contact list.
}
Note that the setting we study here is different from that of AirDrop (or private set intersection generally), as mutual contact discovery for messaging is asynchronous and mediated by a third party (offering the messaging service), whereas AirDrop runs directly and in real-time between the two participants. This difference allows us to implement mutual contact discovery more efficiently, at least for the users whose computation and communication complexity is linear in the length of their own contact list (and does not depend on the length of the contact lists held by the other users). More importantly, mutual contact discovery allows a user to discover its mutual contacts among \emph{all} other users at once.

Finally, we wish to point out the resemblance of mutual contact discovery with contact tracing and exposure notification, recently proposed to fight the COVID-19 pandemic, see~\autocite{who2020contact-tracing,martin2020demystifying-covid19-tracing}. Although they are different in certain fundamental aspects (there is no underlying social graph over pre-existing identifiers, for example), nevertheless techniques used in \eg the DESIRE protocol\footnote{%
  See
  \url{https://github.com/3rd-ways-for-EU-exposure-notification/project-DESIRE}.
}
turned out to be applicable in our setting as well. 

%

\section{Model and notation}
\label{sec-model}

Mutual contact discovery allows a potentially very large, dynamic, set of \term{users} from an existing social graph, that is represented by locally maintained contact lists, to discover each other when joining a new service.
Alice and Bob are mutual contacts if Alice is a contact of Bob and Bob is a contact of Alice. In the case of messaging, the underlying social graph is based on the existing mobile phone network, where the identifiers are mobile phone numbers and the social graph is given by the contact lists users maintain on their mobile phones.

Members can communicate directly, securely, and anonymously with a separate \emph{matching server}, and can communicate with each other only through this matching server.\footnote{%
  This star topology is inherent to (current) messaging services, for which contact discovery is developed. In practice members will have many different ways to find and connect to each other directly, using their unique identifiers, but in the context of this problem we disallow them to use this ability.
}
Intuitively, the mutual contact discovery protocol should satisfy the following requirements.
\begin{description}
\item[Correctness]
  When Alice and Bob are mutual contacts, use the messaging service, are honest, and concurrently run the mutual contact discovery protocol, they will discover each other.
\item[Security]
  If Bob is not a member then Alice cannot be made to believe he is.
\item[Membership privacy]
  If the adversary is not a contact of Alice, then it will not be able to learn whether Alice is a member.
\item[Contact privacy]
  An adversary is not able to tell whether Bob is a contact of Alice, provided Alice and Bob are both honest.
\end{description}
In the above, the adversary could be a user or the matching server itself. We make no assumptions about the behaviour of adversarial users (or the matching server). We do assume that the matching server runs independently of the messaging server, in particular we assume that the matching server has no information about membership. This formal separation of roles makes the analysis in this paper cleaner as it allows us to focus on what exactly a matching server can and cannot learn purely on the basis of the mutual contact discovery protocol alone. (Clearly, the messaging service is able to reconstruct the social graph for those members that actively exchange messages.)

Note that the problem can be made more abstract by considering arbitrary identifiers and arbitrary relationships: all that contact discovery requires is an existing social graph represented by locally maintained contact lists. We now proceed with this more abstract definition.

\subsection{Formal definition}

Assume a directed `social graph' $(\sgu,\sge)$ exists with vertices $\sgu$ representing people, and directed edges $\sge$ representing their relationships.
$A,B,\ldots \in \sgu$ represent the unique identifiers of its members. Vertices know their own identifier, and maintain a list of contacts $\contacts{A} = \{ B \in \sgu \mid (A,B) \in \sge \}$. (Note how we associate a user with its identifier, and how the list of contacts captures the full social graph.) As $\sgu$ may be small, we assume these identifiers have relatively low entropy, and are therefore easily guessed or enumerated.

It turns out that it is hard to prevent a malicious user from evading detection as a mutual contact. We therefore define a slightly more general, weaker, form of mutual contact discovery as follows.  Allow members $M$ to mark certain contacts for whom it wants to hide membership (but for whom it wants to discover whether that contact is a member), and mark the other contacts as visible. That is, define two disjoint sets $\hidden{M}$ and $\visible{M}$ such that $\contacts{M} = \visible{M} \cup \hidden{M}$. For honest members $M$, $\hidden{M} = \emptyset$. Moreover, for \emph{strong} mutual contact discovery, $\hidden{M} = \emptyset$ for \emph{all} members, even malicious ones. (This should then be enforced by the protocol.)

\begin{definition}[mutual contact discovery protocol]
  \label{def-mcd}
A mutual contact discovery protocol among a set of members $\members$ over a social graph $(\sgu,\sge)$ is an interactive protocol run between
\begin{itemize}
\item
  a set of members $\members \subseteq \sgu$, where each member $M \in \members$ has a set of contacts $\contacts{M} \subseteq \sgu$,
  divided into two disjoint sets $\visible{M}$ and $\hidden{M}$,
  and
\item
  a matching server $\matchingserver$.
\end{itemize}
The set of members $\members$ is unknown to both the matching server and each of the members.
The protocol consists of the following two sub-protocols.
\begin{itemize}
\item
  $\mcdsetup(1^\secpar)$ is a probabilistic polynomial-time algorithm with as input a security parameter $1^\secpar$. It generates the necessary system parameters, and initialises all members $M \in \members$ as well as the matching server $\matchingserver$. $\mcdsetup$ must be run first.
\item
  $\mcddiscover(\members)$ is an interactive protocol run in parallel between each of the members $M \in \members$ and the matching server $\matchingserver$. If successful, each $M \in \members$ obtains a set $\out{M}$ of discovered contacts.
\end{itemize}
\end{definition}
Define
\[
\mc{A}{B} \isdefas \mutualcontact{A}{B}~.
\]
Note that this is only a symmetric relationship among honest users, or for \emph{strong} mutual contact discovery.

The protocol must satisfy the following correctness condition.
\begin{definition}[Correctness]
\label{def-mutual}
  For all $A \in \members$ and all $B \in \sgu$ that follow the protocol we have
  \[
    \begin{cases}
       B \in \out{A} & \text{ if  }\> (B \in \members) \wedge \mc{A}{B}, \\
       B \not\in \out{A} & \text{ otherwise, with overwhelming probability.}
    \end{cases}
  \]
\end{definition}
Observe how this abstract definition captures the essence of mutual contact discovery while simplifying from the practical setting of mobile phone users joining a messaging service and discovering their mutual contacts in the following ways.
\begin{itemize}
\item
  The underlying social graph is assumed to be static, while in practice people add and remove contacts from their contact list.
\item
  The set of members $\members$ of the messaging service is assumed to be static as well. Again in practice this is not the case.
\item
  The contact discovery protocol is run once at the same time for all current members in parallel. In practice, members will run the discovery protocol as soon as they join, and occasionally afterwards.
\end{itemize}
As we will discuss later in section~\ref{ssec-reallife}, this does not fundamentally change the problem and the results.

\subsection{System model}

The system is a star network with member devices at the edges and the matching server as the central node. Members can authenticate the server\footnote{%
  In what follows we write server to denote the matching server.
}
and can set up a sender-anonymous secure communication channel that preserves the integrity, confidentiality and freshness of all messages exchanged. In this setting the server is authenticated, while members remain anonymous.\footnote{%
  In the conclusions section we will return to the question of how to keep members anonymous in practice, given that they communicate with the messaging server directly and thus leak their possibly identifying IP address. Note this is a \emph{general} issue that plagues \emph{any} contact discovery protocol that aspires to protect privacy.
}
In fact we assume anonymous communication channels are used throughout (a fresh one for \emph{each message exchange}). Tor\footnote{%
  \url{https://www.torproject.org}
}
could be used for this purpose~\autocite{dingledine2004tor}, or a system like Private Relay recently announced by Apple.\footnote{%
  \url{https://developer.apple.com/videos/play/wwdc2021/10096}
}
We will not go into details here, but simply note that many privacy preserving protocols need to make a similar assumption about the communication layer offering some kind of sender anonymity (or assuming a benevolently amnesiac  server if such anonymity is not offered).

\subsection{Threat model and security assumptions}

We assume an \emph{active} adversary where all adversarial entities may behave arbitrarily. In particular an adversarial member may add extra identifiers to its input $\contacts{X}$, and may decide to hide certain contacts using $\hidden{X}$. Such an adversary can observe, eavesdrop, replay, modify or block messages on the network. Note however that by assumption (see earlier this section) all communication between members and the matching server are secure (confidential and authentic), so in our analysis we only really need to consider observing and blocking messages.

\subsection{Requirements}
\label{ssec-reqs}

We formally define the requirements our protocol should satisfy (as informally defined at the start of this section) using the `malicious model'~\autocite{goldreich1987how-to-play,goldreich2001foundations-crypto-vol2}
(which is weaker than the universal composability framework~\autocite{canetti2001UC}). In this setting, all security, privacy and functional requirements are captured by the description of an \term{ideal system}, where a trusted third party (communicating using secure links) receives all participant inputs, locally computes the results, and returns all participant outputs to each participant individually. This ideal system is defined here as follows.
\begin{definition}[ideal system for mutual contact discovery]
  \label{def-ideal-mcd}
  Let social graph $(\sgu,\sge)$ be given. Let $\members \subseteq \sgu$ be a set of members, and let $\matchingserver$ be the matching server.
  Let $\trustedserver$ be a trusted server connected to each user in $\sgu$ and to the matching server $\matchingserver$  over a secure (and for users anonymous) communication link.

  Let $\hidden{M}$ and $\visible{M}$ be disjoint subsets of $V$ given as input to participant $M$,
  set $\contacts{M}=\hidden{M} \cup \visible{M}$, and let $\err$ be the input to the matching server $\matchingserver$.\footnote{%
    Malicious participants may set their input $\contacts{M}$ to an arbitrary set. For honest  members, $\hidden{M}=\emptyset$. For honest server, $\err=0$. $\err$ models how often the matching server withholds information to participants.
  }
  The ideal system for mutual contact discovery proceeds in the following synchronous phases.
  \begin{enumerate}
  \item
    Each member $M \in \members$ sends $\hidden{M}$ and $\visible{M}$ to trusted server $\trustedserver$.
    The matching server $\matchingserver$ sends $\err$ to trusted server $\trustedserver$.
  \item
    $\trustedserver$ computes, for all $M \in \members$,
    the set $\out{M} = \{ X \mid (X \in \members) \wedge \mc{M}{X} \}$.
    $\trustedserver$ selects, for all $M \in \members$,
    a set $\out{M}' \subseteq \out{M}$ and sets $\err_M = | \out{M} \setminus \out{M}'|$.\footnote{%
      Here $\setminus$ denotes subtraction among sets, and recall that $\mc{M}{X}$ denotes that $M$ and $X$ are mutual contacts.
    }
    The selection of all $\out{M}'$ is free except that $\sum_{M \in \members} \err_M = \err$ must hold.
  \item
    $\trustedserver$ sends $\out{M}'$ to each $M \in \members$ as its results, which each $M$ outputs.
    $\trustedserver$ sends $|\cset|$ and $|\mcset|$ to $\matchingserver$ as its result, which it outputs.\footnote{%
       This output is of course irrelevant in any practical application, but captures what the matching server learns during a protocol run, and by making the output explicit we stay within the general SMC paradigm.
    }
  \end{enumerate}
  Here
  \[
     \cset = \{ (A,B) \in \members \times \members \mid A \in \contacts{B} \}
  \]
  is defined as the set of all contact pairs, 
  and
  \[
    \mcset  =  \{ (A,B) \in \members \times \members \mid \mc{A}{B} \}
  \]
  is defined as the set of all mutual contact pairs.
\end{definition}
Observe how this ideal model also prevents the matching server form learning the individual number of contact held by each member, a privacy property currently not achieved by other contact discovery protocols.

For ease of exposition we do not consider aborts; those are modelled as participants not submitting their full set of contacts. However, the definition of the ideal system is somewhat tricky because the matching server is an active participant of the real system, that may also misbehave. In particular we have to allow it to be only partially responsive, omitting to relay certain information to certain participants. This is what the error input $\err$ is for.

The behaviour of the \term{real protocol} must subsequently be proven to be indistinguishable from the behaviour of the ideal model when simulated by a simulator corrupting the same set of parties involved. This is proven using a simulation proof technique~\autocite{lindell2017how-to-simulate}.
\begin{definition}[secure mutual contact discovery protocol]
  \label{def-secmcdp}
  A protocol $\prot$ (satisfying the syntactic of definition~\ref{def-mcd} and correctness constraints of definition~\ref{def-mutual}) is a secure mutual contact discovery protocol if for every adversary $\adversary{\prot}$ (not tapping any communication channels) there exists a simulator $\simulator{\ideal}$ for the ideal system $\ideal$ of definition~\ref{def-ideal-mcd}, where $\simulator{\ideal}$ controls the same set of entities as $\adversary{\prot}$, such that the joint view of $(\prot,\adversary{\prot})$ is indistinguishable from
  the joint view of $(\ideal,\simulator{\ideal})$.
\end{definition}
We define strong mutual contact discovery as follows.
\begin{definition}[secure strong mutual contact discovery protocol]
  A protocol $\prot$ is a secure strong mutual contact discovery protocol if it is a secure mutual contact discovery protocol according to definition~\ref{def-secmcdp} while for \emph{all} members $M$,
  $\hidden{M} = \emptyset$ in the ideal system.
\end{definition}

\subsection{Pairings and the BDH assumption}

Let $q$ be a large prime. Let $\G{1}$ and $\G{2}$ be two groups of order $q$. Let $\pairingsym: \G{1} \times \G{1} \mapsto \G{2}$ be a \term{bilinear map} (also known as \term{pairing}) satisfying the following properties:
\begin{itemize}
\item
  For all $a,b \in \Z$ and all $P,Q \in \G{1}$ we have
  $\pairing{aP}{bQ} = \pairing{P}{Q}^{ab}$.
\item
  $\pairingsym$ is non-degenerate: for some $P,Q \in \G{1}$ the pairing $\pairing{P}{Q}$ is unequal to the identity element of $\G{2}$.
\item
  $\pairingsym$ can be efficiently computed.
\end{itemize}
A pairing satisfying these requirements is called \term{admissible}. Such groups and an admissible pairing exist~\autocite{boneh2003idcrypto}.
\begin{definition}[Bilinear Diffie-Hellman (BDH) problem]
  The Bilinear Diffie-Hellman (BDH) problem over groups $\G{1}, \G{2}$ of order $q$, and an admissible pairing $\pairingsym: \G{1} \times \G{1} \mapsto \G{2}$ is defined as follows: given a random generator $P$ of $\G{1}$ and
  $aP, bP, cP \in \G{1}$ for some random $a,b,c \in \Z_q^*$, compute $\pairing{P}{P}^{abc}$. 
\end{definition}
It is believed that the Bilinear Diffie-Hellman problem is hard (\ie any randomised polynomial time adversary has at most a negligible advantage solving BDH) for certain choices of $\G{1}$, $\G{2}$, and $\pairingsym$~\autocite{boneh2003idcrypto}.

\section{A protocol using certified identifiers}
\label{sec-protocol}

Our protocol resembles Boneh and Franklin's system for Identity Based Encryption~\autocite{boneh2003idcrypto}). Let $\H{1}:\sgu \mapsto \G{1}$ be a cryptographic hash function that maps identities to points in $\G{1}$. We define
\[
  \point{A} = \H{1}(A)~.
\]
Define the certificate $\cert{A}$ of a member identity $A$ to be
\[
\cert{A} = s \point{A}
\]
for some secret $s \in \Z_q^*$. Furthermore define a \emph{token} for a pair of users $A, B \in \sgu$ as
\[
\token{A}{B} = \pairing{\cert{A}}{\point{B}}~.
\]
We have the following property
\begin{property}
  \label{prop-eqtoken}
  For all $A,B \in \sgu$ we have $\token{A}{B} = \token{B}{A}$.
\end{property}
\begin{proof}
  As $P$ is a generator of $\G{1}$ there are $a,b \in \Z_q^*$ such that $\point{A} = aP$ and $\point{B}=bP$. Then $\token{A}{B} = \pairing{saP}{bP}
  = \pairing{P}{P}^{sap} = \pairing{sbP}{aP} = \token{B}{A}$.
  \qed
\end{proof}

The intuition of the next protocol is that the above property allows mutual contacts to both compute the same token to discover each other, but as the computation of $\token{A}{B}$ requires an unforgeable (and secret) identifier certificate as input, no one else can compute this token (this is formally proven later in theorem~\ref{th-bdh-mcdt}).

The protocol runs in two phases. In the \emph{submission phase}, each member $M$ essentially submits a different token $\token{M}{A}$ to the matching server for each of its contacts $A$ (using an additional hash function to shield the actual token from that server). The matching server stores all hashed tokens received. If $A$ also participates and has $M$ as contact, the server will have two copies of that hashed token (recall that $\token{M}{A} = \token{A}{M}$) in its database at the end of the submission phase. In the \emph{query phase}, $M$ asks the server how many copies of the token $\token{M}{A}$ it holds. If the server response indicates there are two such copies, a mutual contact is found.
Tokens are hashed before submitting them to the server to prevent the server from constructing a valid response to a query, and thus preventing it from lying about a mutual contact that does not in fact exist.

The full protocol is specified as follows.
\begin{definition}[Mutual contact discovery protocol]
  \label{def-prot-mcdp}
  Let $\point{A}$, $\cert{A}$ and $\token{A}{B}$ be as defined above. Define the following protocol.
\begin{itemize}
\item
  The setup phase $\mcdsetup(1^\secpar)$ generates a large prime $q$ (of length $\secpar$), groups $\G{1}$ and $\G{2}$ of order $q$, an admissible pairing $\pairingsym: \G{1} \times \G{1} \mapsto \G{2}$, a generator $P$ of $\G{1}$, a secret $s \in \Z_q^*$, a public key $\Ppub = sP$,
  and cryptographic hash functions $\H{1}:\sgu \mapsto \G{1} \setminus P$ and $\H{2}: \G{2} \times \G{1} \times \G{1} \mapsto \{0,1\}^n$ (for some choice of $n$, also of length $\secpar$).

  Let $<$ be a global order over $\G{1}$, and define
  \[
  \HT(X,Y,Z) =
    \begin{cases}
    \H{2}(X,Y,Z) & \text{ if  } Y<Z \\
    \H{2}(X,Z,Y) & \text{ otherwise. }
    \end{cases}
  \]
  Note that $\HT(X,Y,Z)=\HT(X,Z,Y)$.
  
  The setup phase publishes
  $\{ q,\G{1},\G{2},\pairingsym,P,\Ppub,\H{1},\H{2} \}$ and provides each member $M \in \members$ its certificate
  $\cert{M} = s \point{M} = s \H{1}(M)$. Members keep this certificate secret. The setup phase instructs the matching server $\matchingserver$ to initialise a set $S = \emptyset$. (To be clear, the matching server does not learn about the secret values.)
  $s$ is erased at the end of the setup phase.\footnote{%
    Recall that we are currently only considering the static, one shot, setting as defined in definition~\ref{def-mcd}, where an abstract trusted entity runs the setup phase and provides members their certificates.   
    We will return to this in section~\ref{ssec-certs}.
  }
  
\item
  The interactive discovery protocol $\mcddiscover(\members)$ between a member $M \in \members$ and the matching server $\matchingserver$ runs in two phases, as follows.
  \begin{itemize}
  \item
    In the \emph{submission phase} each member $M$, for each $A \in \contacts{M}$, computes token $\token{M}{A}$. $M$ then computes \emph{augmented} tokens $\tokenx{M}{A}{M}{A}$ and $\tokenx{M}{A}{A}{A}$. It sends
    $\tuple{\tokenx{M}{A}{M}{A},\tokenx{M}{A}{A}{A}}$ to the matching server, which stores it in $S$.
  \item
    In the \emph{query phase} each member $M$ does the same and again sends the tuple
    $\tuple{\tokenx{M}{A}{M}{A},\tokenx{M}{A}{A}{A}}$ to the matching server.
    The matching server returns all tuples in $S$ that match $\tokenx{M}{A}{M}{A}$ on the first component but are different on the second component.\footnote{%
      If an external observer should not be allowed to notice whether a query returns a mutual contact or not, the response should be made to be exactly one (possibly random) tuple. To simplify the proof we assume that whatever the matching server learns is public knowledge.
    }
    $M$ then computes $\tokenx{M}{A}{M}{M}$ and verifies whether at least one of the tuples returned has this value as the second component. If so, $M$ records that $A$ is discovered as a mutual contact, adding $A$ to $\out{M}$.
  \item
    After the query phase, each $M \in \members$ outputs $\out{M}$. The matching server outputs 
    $| \{ (X,Y),(X',Y') \in S \mid X=X' \wedge Y \neq Y'\} |$ as value for $|\mcset|$ and $|S|$ as value for $|\cset|$.
  \end{itemize}
\end{itemize}
\end{definition}
\begin{observation}
  \label{obs-tsym}
  We have $\tokenx{M}{A}{M}{A} = \tokenx{A}{M}{A}{M}$, so if $A$ and $M$ are mutual contacts and are both honest, they will discover each other (provided the matching server is honest as well).
\end{observation}
This follows from property~\ref{prop-eqtoken} and the definition of $\HT$.

Note that by forcing the server to respond with $\tokenx{M}{A}{M}{M}$ in the query phase ensures that it cannot trick $M$ into believing $A$ is a member when in fact it is not: only $A$ or $M$ could have created that message (because $\token{M}{A}$ is never revealed in the clear).

Recall that we assume communication links to be sender anonymous, and observe that indeed the protocol in no way relies on the matching server to be able to link different messages from the same user. 

\begin{observation}
\label{obs-csym}
When all participants in a run of protocol~\ref{def-prot-mcdp} are honest, then $| \{ (X,Y),(X',Y') \in S \mid X=X' \wedge Y \neq Y'\} |$ and
$|S|$ as returned by the protocol for $|\mcset|$ and $|\cset|$ are indeed the correct values according to definition~\ref{def-ideal-mcd}.
\end{observation}
This follows from the fact that every honest member submits exactly one token pair for every contact, and that participants that are mutual contacts submit a token pair with the same first component (by observation~\ref{obs-tsym}).

\subsection{Analysis}

Model the hash functions $\H{1},\H{2}$ as random oracles~\autocite{bellare1993randomoracles,katz2014modern-crypto}. Assume the Bilinear Diffie-Hellman problem is hard for our choice of $\G{1}$, $\G{2}$, and $\pairingsym$. Security of our protocol depends on the hardness of the following problem in this setting.
\begin{definition}[Mutual Contact Discovery Token (MCDT) problem]
  The Mutual Contact Discovery Token (MCDT) problem over 
  groups $\G{1}, \G{2}$ of order $q$, an admissible pairing $\pairingsym: \G{1} \times \G{1} \mapsto \G{2}$, a generator $P$ of $\G{1}$, a public key $\Ppub = sP$ (for some random secret $s \in \Z_q^*$), and cryptographic hash functions $\H{1}:\sgu \mapsto \G{1}$ is defined as follows: given the certificates $C(X)=s\point{X}$ for some $X \in \mathcal{V} \subset \sgu$ chosen by the adversary, and
  random $A,B \in \sgu$ with $ A \neq B$ and $A,B \notin \mathcal{V}$, compute $\token{A}{B} = \pairing{\cert{A}}{\point{B}}$ (where
  $\cert{X}=s\point{X}$ and $\point{X}=\H{1}(X)$ for $X \in \sgu$).
\end{definition}
\begin{theorem}
  \label{th-bdh-mcdt}
  If the Bilinear Diffie-Hellman (BDH) problem over groups $\G{1}, \G{2}$ of order $q$, and an admissible pairing $\pairingsym: \G{1} \times \G{1} \mapsto \G{2}$ is hard, then so is the Mutual Contact Discovery Token (MCDT) problem
  over the same $\G{1},\G{2},\pairingsym$, a generator $P$ of $\G{1}$, a public key $P_{\mathrm{pub}} = sP$ and cryptographic hash functions $\H{1}:\sgu \mapsto \G{1}$ (in the programmable random oracle model).
\end{theorem}
\begin{proof}
  Suppose not.
  
  Let $\adversary{MCDT}$ be an adversary for the MCDT problem. We show how to convert it into an adversary $\adversary{BDH}$ for the BDH problem as follows, where $\adversary{BDH}$ functions as the challenger for $\adversary{MCDT}$.

  Let the setup of $\adversary{BDH}$ be groups $\G{1}, \G{2}$ of order $q$, and an admissible pairing $\pairingsym: \G{1} \times \G{1} \mapsto \G{2}$.

  Let $P$ be a generator of $\G{1}$, and let $P_1 = sP$, $P_2 = aP$, and $P_3 = bP$ all in $\G{1}$ for some random $s, a, b \in \Z_q^*$ be the challenge given to $\adversary{BDH}$.

  $\adversary{BDH}$ sets up $\adversary{MCDT}$ as follows: $\G{1}, \G{2}$, $q$, and $\pairingsym$ as given, $P$ as in the challenge, and $\Ppub = P_1$, also from the challenge. $\H{1}$ is a random oracle controlled by $\adversary{BDH}$ as described below, that can be queried by $\adversary{MCDT}$.

  \renewcommand{\L}{L_1}
  
  $\adversary{BDH}$ answers queries to $\H{1}$ as follows. It maintains a mapping $\L$ of already made queries and their responses. $\L$ is initially empty. When queried for a point $X \in \G{1}$ not in $\L$ it picks a random $b \in \Z_q^*$, returns $bP$, and stores $\L[X]=(b,bP)$. If $X$ is in $\L$ it looks up $(b,Q) = \L[X]$ and returns $Q$. Observe how this guarantees a random distribution for the output of the $\H{1}$ oracle.
  
  Next $\adversary{BDH}$ allows $\adversary{MCDT}$ to query the certificates
  $C(X)=s\point{X}$ for some $X \in \mathcal{V} \subset \sgu$. $\adversary{BDH}$ responds by first (internally) querying the oracle for each $X$ and subsequently looking up $\L[X]$. This equals $(b,bP)$ for some $b$ by construction. It responds by returning $b \Ppub = b sP = s bP = s \H{1}(X) = s \point{X}$ by construction.

  Then $\adversary{MCDT}$ starts and is allowed queries to the $\H{1}$ oracle. Let us assume it makes $q_E$ such queries.

  $\adversary{BDH}$ then sets up the challenge $A$, $B$ to $\adversary{MCDT}$ as follows: it picks random $A$ and $B$ and looks up $A$ and $B$ in $\L$. If any of these exists in $\L$, the adversary fails. Observe that the adversary is successful with non-negligible probability at least $(1- q_E/|\sgu|)^2$. (The queries made by 
  $\adversary{BDH}$ to compute the certificates do not count as by assumption $A,B \notin \mathcal{V}$.) Otherwise it returns $P_2$ for $\H{1}(A)=\point{A}$ and $P_3$ for $\H{1}(B)=\point{B}$ and updates $\L$ accordingly (the random $b$ is no longer relevant and can be kept empty). It then sends $A$ and $B$ as the challenge to $\adversary{MCDT}$.

  $\adversary{MCDT}$ runs and is allowed additional queries to the $\H{1}$ oracle.
  
  After some time $\adversary{MCDT}$ responds with some non-negligible probability $p$ with the correct response $\token{A}{B} = \pairing{\cert{A}}{\point{B}}$. But then
  $\token{A}{B} = \pairing{s\point{A}}{\point{B}} = \pairing{sP_2}{P_3} = \pairing{saP}{bP}=\pairing{P}{P}^{sab}$. In other words $\token{A}{B}$ is a correct response to the BDH challenge, which implies $\adversary{BDH}$ wins with non-negligible probability $p(1- q_E/|\sgu|)^2$, contrary to assumption.
\qed  
\end{proof}
We now prove that any view of an adversarial entity in a real protocol execution can be simulated by a simulator in the ideal system, as required by definition~\ref{def-secmcdp}.

\begin{theorem}
  The mutual contact discovery protocol~\ref{def-prot-mcdp} is secure according to definition~\ref{def-secmcdp}; it's behaviour cannot be distinguished from the ideal system~\ref{def-ideal-mcd}.
\end{theorem}
\begin{proof}
Informally, the theorem is based on the following reasoning. Theorem~\ref{th-bdh-mcdt} ensures that only $A$ and $M$ can compute a token $\token{M}{A}$. As $\token{M}{A}$ is only ever sent encapsulated by $\H{2}$, which is modelled by a random oracle, no other party learns any information about $\token{M}{A}$. We conclude that only $A$ and $M$ can compute an augmented token $\tokenx{M}{A}{X}{Y}$ for any $X$ and $Y$, but observe that the matching server $\matchingserver$ learns such values when they are sent (by either $A$ or $M$ as just argued). Also note that by the same reasoning a valid tuple $\tuple{\tokenx{M}{A}{M}{A},\tokenx{M}{A}{A}{A}}$ (\ie a tuple where the first and second part actually belong together) can only be constructed by $A$ or $M$ (you cannot even see from the outside whether $\tokenx{M}{A}{X}{Y}$ and $\tokenx{A'}{M'}{X'}{Y'}$ match on $A=A'$ and $M=M'$). This means we only need to consider the originator of such tuples: a malicious server could share such tuples with other users but this would in the end only result in the server adding meaningful, valid, tuples to its set $S$ that it already has. Note that invalid tuples (that either combine otherwise valid hash values but corresponding to different tokens, or combine random values) are ignored in the query phase by the querying user.

We now proceed with a more formal, detailed, proof.

\renewcommand{\L}{L_2}

In the following, the simulator provides the adversary oracle access to the hash function $\H{2}$, maintaining a mapping of already returned responses $\L$, initially empty. Whenever the adversary queries the oracle for an input $(X,Y,Z)$ the simulator looks up $\L[(X,Y,Z)]$ and returns that as the response if it exists, or creates a random element from $\{0,1\}^n$, stores it in $\L[(X,Y,Z)]$ and returns that. $\L$ is maintained such that given an earlier response $x$ the simulator can efficiently retrieve $(X,Y,Z)$ such that $\L[(X,Y,Z)]=x$

\proofpart{Simulating a participant $M$}

We start with the submission phase. The simulator is given $\cert{M}$ and starts adversarial $M$. $M$ will return a list of tuples to submit to the matching server. The task of the simulator is to extract the input $\hidden{M}'$, $\visible{M}'$ and hence $\contacts{M}'$. It can do so as follows, considering the tuples one by one.

Let $\tuple{H',H''}$ be the next tuple. According to the protocol, this tuple should match $\tuple{\tokenx{M}{A}{M}{A},\tokenx{M}{A}{A}{A}}$ for some $A$. The simulator looks up the preimage of $H'$ and $H''$ in $\L$.

If the preimage of $H'$ does not exist, it skips this tuple and considers the next. (It can do so because then, with overwhelming probability, no other honest member $A'$ will submit a valid token $\tokenx{A'}{M}{A'}{M}=H'$ that could result in a match at the server.)

Otherwise, let $(T,X,Y)$ be this preimage, \ie it is the result of calling $\H{2}$ for either
$\HT(T,X,Y)=H'$ or $\HT(T,Y,X)=H'$ . The simulator verifies that $X<Y$. If not,
it skips this tuple and considers the next. (It can do so because no other honest member $A'$ will query the oracle with $X \ge Y$.) The simulator then verifies that $X=M$ and
$T = \token{X}{Y} = \pairing{\cert{M}}{Y}$ and sets $A'=Y$ (or that $Y=M$ and
$T = \token{Y}{X} = \pairing{\cert{M}}{X}$ and sets $A'=X$).
If this verification step fails, then it also skips this tuple and considers the next. (It can do so because then, with overwhelming probability, again no other honest member $A'$ will submit a valid tuple with $H'$ as the first field. This is because according to theorem~\ref{th-bdh-mcdt}, $M$ can only construct a valid token $T$ of the form $\pairing{\cert{M}}{Z}$ for some $Z$, and by the construction of a valid $H'$, $Z$ should equal either $X$ or $Y$.) 

If the preimage of $H''$ does not exist, then the simulator adds $A'$ to $\hidden{M}'$ and proceeds with the next tuple. (Because then, with overwhelming probability, $A'$ will not detect $M$ as a mutual contact, while $M$ will detect $A'$ if it is a mutual contact and participating.)

Otherwise, let $(T',X',Y')$ be this preimage, \ie $\HT(T',X',Y')=H''$. The simulator then verifies that $T=T'$ and $X'=Y'=A'$. If any of these tests fail, it adds  $A'$ to $\hidden{M}'$ (again because in this case $A'$ will not detect $M$ as a contact). Otherwise it adds $A'$ to $\visible{M}'$.  It then proceeds with the next tuple.

When all tuples have been considered, the simulator removes any $A'$ from $\hidden{M}'$ that are also a member of $\visible{M}'$.

We now consider the query phase. 
\begin{enumerate}
\item
  The simulator receives $\out{M}$ from the trusted third party.
\item
  For every $\tuple{X,Y}$ it receives from the adversary the simulator returns all tuples stored locally during the submission phase for which $X$ matches the first component and $Y$ does not match the second component. (Observe how this is corresponds exactly to what the matching server would return if interacting with the adversarial $M$ based on what it received from $M$ itself. This ensures the adversary cannot distinguish interacting with the simulator or the real server.) The simulator also checks whether
  $X = \tokenx{M}{A}{M}{A}$ for some $A \in \contacts{M}$ (again it can do so). If this is not the case, nothing happens. Otherwise, if $A \in \out{M}$ it adds $(\tokenx{M}{A}{M}{A},\tokenx{M}{A}{M}{M})$ to its response to the adversary (if it hadn't added that tuple already). 
\item
 The simulator outputs whatever the adversary outputs.
\end{enumerate}
With the above observations in mind, it is easy to see that the view of the adversary in the real protocol execution is indistinguishable from its view provided by the simulator. Therefore its output is the same, and hence the output of the simulator in the ideal case equals the output in the real case.

\proofpart{Simulating the matching server}

We start with the submission phase. The simulator receives $\hidden{M}'$ and $\visible{M}'$ from all $M$ and uses that to compute $|\cset|$ and $|\mcset|$.
It interacts with the (adversarial) matching server as follows.

We use observation~\ref{obs-csym} to construct the necessary information that the simulator needs to send the right (number of) messages in the submission and the query phase.
Recall these messages are pairs $(X,Y)$ of augmented tags.\footnote{%
  And by assumption do not contain any information about its sender, even in the lower layer.
}
We know that $|\cset|$ equals $|S|$ and hence denotes the number of submission messages the simulator needs to send. Of those messages, $|\mcset|$ denotes the number of times a particular augmented tag occurs twice as $X$ (the first element) in such pairs. 

Randomly generate $|\mcset|$ `tokens' $T_i$ and points $P_i$ and $P'_i$ and compute $X_i = \HT(T_i,P_i,P'_i)$, $Y_i = \HT(T_i,P_i,P_i)$, and $Y'_i = \HT(T_i,P'_i,P'_i)$. Send $\tuple{X_i,Y_i}$ and $\tuple{X_i,Y'_i}$ to the matching server. The random oracle for $\H{2}$ ensures that these are indistinguishable from the actual pairs of tuples the matching server receives for actual mutual contacts in the real execution of the protocol.

For the remaining $|\cset| - 2|\mcset|$ contacts, generate random `tokens' $T'_i$ and points $P''_i$, compute $X'_i = \HT(T'_i,P_i,P''_i)$ and $Y''_i = \HT(T_i,P''_i,P''_i)$, and send $\tuple{X'_i,Y''_i}$ to the matching server. The random oracle for $\H{2}$ again ensures that these are indistinguishable from the tuples the matching server receives for actual non-mutual contacts in the real execution of the protocol.

We now turn to the query phase. For each of the $X_i = \HT(T_i,P_i,P'_i)$ computed for mutual contacts in the submission phase, the simulator sends the queries
$\tuple{X_i,Y_i}$ and $\tuple{X_i,Y'_i}$ to the server.
For each of the $X'_i = \HT(T'_i,P_i,P''_i)$ computed for non-mutual contacts in the submission phase, it sends the query $\tuple{X'_i,Y''_i}$. These messages correspond exactly to the messages the real (honest) participants would send.

The simulator initialises $\err$ to $0$.

Given that the simulator sent all submissions to the server itself, it knows exactly what responses to expect. In particular, for each query
$\tuple{X_i,Y_i}$ corresponding to a \emph{mutual} contact, it expects the response $\tuple{X_i,Y'_i}$ (and vice versa), while for a query
$\tuple{X'_i,Y''_i}$ corresponding to a \emph{non-mutual} contact it expects no response at all. If it doesn't receive a response, or receives and invalid response when querying for a (internally known) mutual contact, it increases $\err$ by one.

At the end of the query phase, the simulator sends $\err$ to the trusted server. $\err$ exactly corresponds to the number of times a participant fails to discover a mutual contact because the matching server lies or remains silent when queried. The resulting output in the ideal system is therefore indistinguishable from the output that would have resulted if the adversarial server had interacted with the real world.
\qed
\end{proof}

\subsubsection{Preventing denial of service attacks}

Participants are not authenticated (for obvious privacy reasons), and the number of submissions or queries that can be submitted is not limited in any way. As a result anybody can flood the matching server with requests and overload its database $S$. Such denial of service attacks can be prevented as follows.
\begin{itemize}
\item
  Members also receive a \emph{credential}\footnote{
    For example an Idemix like attribute based credential, see~\autocite{camenisch2001anoncred,idemix2012}
  }
  $\cred{M}$ over their identity $M$ during the setup phase (in practice provided by the underlying social graph) alongside their certificate $\cert{M}$. Members need to prove ownership of such a credential anonymously in order to be able to communicate with the matching server. This blocks non-members mounting a denial of service attack.
\item
  To prevent members from flooding the matching server with requests, some kind of rate limiting needs to be implemented. This could be based on proof-of-work~\autocite{back1997hashcash,dwork1993pricing}, requiring participants to solve a moderately hard puzzle each time the want to connect to the matching server. Or a maximum $n$ on the number of contacts in any lists could be assumed, and members are issued $n$-times anonymous credentials~\autocite{camenisch2006clone-wars} that can be used at most $n$ times to contact the matching server.
\end{itemize}

\subsubsection{Network based inference attacks}

Note that the size of the individual member's list of mutual contacts is only hidden when members submit and query their contacts one at a time, as our proposed protocol indeed does. Also note that the protocol does \emph{not} prevent side channel attacks where the server assumes subsequent requests that happen in a short frame of time are all coming from the same user. In fact, this type of attack is out of scope of the ideal model, as in the UC (universal composability) models for mix networks~\autocite{wilkstrom2004uc-mix} and onion routing~\autocite{camenisch2005formal-onion}. As~\autocite{camenisch2005formal-onion} explains, this is necessarily so:
\begin{quote}
  [I]t is known how to break security of mix networks using statistics on network usage where the amount of traffic sent and received by each party is not prescribed to be equal, but rather there is a continuous flow of traffic. In cryptography, however, this is a classical situation. For example, semantic security was introduced to capture what the adversary already knows
  about the plaintext (before the ciphertext is even formed) by requiring that a cryptosystem be secure for all a-priori distributions on the plaintext, even those chosen by the adversary. Thus, the cryptographic issue of secure encryption, was separated from the non-cryptographic modelling of the adversary’s a-priori information. [\ldots]
  
  An onion routing scheme can provide some amount of anonymity when a message is sent through a sufficient number of honest onion routers and there is enough traffic on the network overall. However, nothing can really be inferred about how much anonymity an onion routing algorithm provides without a model that captures network traffic appropriately. As a result, security must be defined with the view of ensuring that the cryptographic aspects of a solution remain secure even in the worst-case network scenario.
\end{quote}

\subsection{The dynamic, real life setting}
\label{ssec-reallife}

The protocol presented satisfies the static setting of a mutual contact discovery protocol according to definition~\ref{def-mcd}. As already mentioned earlier, in practice the setting is much more dynamic: the underlying social graph is not static, and people join and leave at arbitrary times. Therefore, in practice the protocol should allow members to repeatedly check for new contacts. Also, what happens when people join and leave deserves more attention.

A simple modification allows the protocol to also cater for this dynamic, real life setting as follows. Instead of running synchronously and distinguishing a submission and query phase, the protocol now runs asynchronously forever and only supports querying. Initially the token database $S$ of the server is empty.
Members $M$ can query for contacts $A$ at arbitrary times, using the same tuple $\tuple{\tokenx{M}{A}{M}{A},\tokenx{M}{A}{A}{A}}$ as in the original protocol. Whenever the matching server receives a new tuple, it adds it to its database $S$. Also, as a response to any such query, the matching server returns all tuples in $S$ that match $\tokenx{M}{A}{M}{A}$ on the first component but are different on the second component.

Because the protocol is now asynchronous, if $M$ queries for contact $A$ before $A$ queries for contact $M$, only $A$ will be informed of a mutual contact. This means that in practice if a member adds a new contact to their contact list, they should repeatedly query the matching server for this new contact until a match is found. Contacts for which mutual contact has already been established do in principle not need to be queried again.

However, this approach only considers the case where contacts are added. What about the case where members delete contacts from their list? For this the protocol could be amended to allow users to send a delete message to the matching server. (Recall from the discussion in the introduction that users do not necessarily always want to inform a contact that they were deleted from their contact list.) To delete a contact $A$, member $M$ sends the tuple $\tuple{\tokenx{M}{A}{M}{A},\tokenx{M}{A}{A}{A}}$, asking the matching server to delete it from $S$. If we trust the server to honour such requests, this ensures that if $A$ later joins as a member, $M$ will not be discovered as a mutual contact by $A$ (and $M$ will also not be informed of $A$'s presence). However, if $A$ already was a member, it would have discovered $M$ as a mutual contact earlier. To prune their mutual contact list, members will have to check the status of \emph{all} their contacts once in a while, by querying the matching server.

We argue that these modifications do not alter the privacy properties in any significant way. The server will only receive
$\tuple{\tokenx{M}{A}{M}{A},\tokenx{M}{A}{A}{A}}$ messages for any $A$ that was in $\contacts{M}$ at any point in time. Those are the same messages it would have received when running the static protocol in a state where \emph{all} these $A$ are a contact of $M$ at the same time. Also, as far as the matching server is concerned all these messages are completely unrelated (because different hashed tokens cannot be linked) and could have come form many different members instead. The additional timing information about when particular messages are sent thus becomes essentially useless. We conclude the matching server does not learn anything more significant than it could have learned from running the static protocol once. Any member will similarly receive only 
tuples $\tuple{\tokenx{M}{A}{M}{A},\tokenx{A}{M}{A}{A}}$ for any $A$ that was a mutual contact at any point in time, which it would have received when running the static protocol once.

\subsection{Where do the certificates come from?}
\label{ssec-certs}

Another fundamental question is of course where the certificates used come from. One way to look at this is by considering certificates being provided by the existing, underlying, social graph. One setup that could work in practice would be one where we leverage the existing infrastructure of mobile phones (messaging services typically use mobile phone numbers as identifiers) and let the mobile network operator (that hands out these mobile phone numbers to subscribers) issue the certificates.\footnote{%
  Whether it is realistic to believe that mobile network operators will do so in order to cooperate with, e.g., Signal is debatable, but the point here is to show that alternative structures to make use of an existing social graph are possible. 
}
This would have the additional advantage that this prevents users from getting certificates for phone numbers that are not their own, as the mobile network operator can strongly authenticate its users through the SIM card they issue to their users.

A separate certificate issuer is also an option, but note that $s$ must be kept secret and should not be managed by the matching server. A natural approach would be to host it along the messaging server, as this is already assumed not to collude with the matching server and by definition already learns the set of members. The method used by the messaging server to authenticate the identity of new users could be used to prevent users from obtaining certificates for identities that do not belong to them. The best approach would be to implement this in trusted hardware (as Signal does now for the \emph{full contact discovery process}), to prevent the secret $s$ from being leaked.

\subsection{Other considerations}
\label{ssec-other}

A subtle issue is the separation between the matching server and the messaging server. In particular, the question is to what extent a malicious user can bypass \emph{mutual} contact discovery. Modern end-to-end encrypted messaging services implement a public key directory containing the keys for their members, so a malicious user could in principle directly ask (through the messaging API) the messaging server for a key of a contact to detect membership. The UI of the messaging app typically prevents this, but the underlying APIs would still allow this.

A more secure implementation of mutual contact discovery therefore needs to prevent this at the API layer, by providing members an access token for each successfully discovered mutual contact. This access token is required to obtain the public key of a contact from the public key directory (which would otherwise refuse to respond, thus protecting knowledge of membership form non-mutual contacts). 

A straightforward idea is to reuse the token already returned by the matching server when $M$ queries it to detect whether $A$ is a mutual contact or not. Recall that in case of $A$ being a mutual contact, the  matching server returns
$\tokenx{A}{M}{M}{M}$. If $A$ also submits 
$\tokenx{A}{M}{M}{M}$ to the messaging server as an access condition to access its public key when adding $M$ to its contact list, then this token can be used by $M$ to obtain $A$'s public key, but only when $A$ and $M$ are mutual contacts. (Here we use the fact that the matching server and the messaging server do not collude.)

\section{Conclusions and discussion}
\label{sec-concl}

Contact discovery is based on the observation that ``access to an existing social graph makes building social apps much easier''~\autocite{moxie2014contact-discovery}. We have shown that by levering such a social graph as the source for certified identifiers, mutual contact discovery can securely and privately be implemented in a malicious adversarial setting.

Our protocol only provides \emph{weak} mutual contact discovery, and allows a malicious mutual contact to evade discovery. A protocol for strong mutual contact discovery would require the matching server to verify that every tuple submitted to the matching server is correct (without learning about its contents), in the sense that it is guaranteed that a potential mutual contact will also discover the member submitting this tuple. This would probably require zero-knowledge proofs of some form, which would be prohibitively expensive. This is left for further research.

We did not implement a prototype and therefore are unable to offer true benchmarks. But let us consider the protocol for the dynamic, real life setting
explained in section~\ref{ssec-reallife}, and argue why this could be practical.

First consider computations. Note how the server is not involved in any complicated computations: it simply stores and looks up values, and answers queries. From the server side this is as good as it could possibly get, in terms of computations. Clients need to compute a pairing though, which may be expensive, but they only need to do so \emph{once} for each member they add to their contact list, if they store the resulting token for later re-use.

Now consider storage. Server storage is a concern. Given the real-life numbers quoted earlier in section~\ref{sec-sota} -- a quarter billion registered users with a thousand contacts each -- implies a database containing dozens of terabytes (instead of the dozens of gigabytes for registered phone numbers otherwise needed to implement straightforward contact discovery, like Signal does~\cite{moxie2017contact-discovery}). In other words, the database grows by a factor corresponding to the average size of the contact lists. Terabyte-size high-traffic databases are used in practice however,\footnote{%
  See e.g. \url{https://www.adyen.com/blog/updating-a-50-terabyte-postgresql-database}.
}
and in this case only need to support a lookup using the primary key (the first hashed token in the tuple). The database can even be sharded, based on this primary key, to distribute load.

Finally consider network traffic and related to that server. Clients can essentially run mutual contact discovery as in the traditional one-sided case, querying the server when a contact is added and regularly refreshing the state for already discovered contacts. Deletion is a new operation, but will be at most as frequent as adding a contact. We see that the number of requests to the server (determining network traffic and server load) are essentially the same as in the traditional one-sided case currently being used on a global scale by apps like WhatsApp and Signal. And given the protocol, the matching server does not have to be implemented in trusted hardware in this case.

We conclude that it should be possible to efficiently implement the proposed protocol in practice, with performance comparable to traditional one-sided contact discovery.


Subtle privacy issues remain. For example, if $B$ knows $A$ is using the service and $A \in \contacts{B}$, then if $B$ doesn't get $A \in \out{B}$ then it learns $A$ didn't put $B$ on their contact list.

In our analysis we have somewhat artificially separated the matching server from the main messaging service, to clearly delineate what can be learned \emph{from the matching protocol alone}. Clearly membership privacy against the server is not really something that can be required in any meaningful sense in practice: the service provider needs to know its members.

Mutual contact discovery restricts the notification of new contacts to those that are mutual. As explained in section~\ref{ssec-other} the messaging services themselves should (and can) technically prevent non-mutual contacts to obtain a key or send a message. This will prevent (malicious) users to discover members automatically using some form of scripting.


The full benefit of privacy friendly contact discovery (one-sided, or mutual) is of course only achieved when combined with truly anonymous messaging services~\autocite{hoepman2015bulletinboard,corrigan-gibbs2015riposte,hooff2015vuvuzela} so that mutual contacts cannot be discovered even by the messaging service itself while delivering the messages. But these services typically do not rely on a small entropy identifier like a phone number to identify contacts, but on public keys instead. This makes the problem inherently easier to solve. It is often left out of scope to discuss how members obtain the public keys of other members. But if members need to meet anyway to exchange public keys, this of course establishes mutual contact by definition, and the problem evaporates all together.  

%

%% file: mutual-contact-discovery-appendix.tex
\section{Introduction to some other approaches}

The protocol presented in the main body of this paper has the best privacy and security properties of all protocols we considered for solving mutual contact discovery. However, the use of pairings may make this a less desirable protocol in practice. It is therefore interesting to also briefly discuss some of the other ideas we had, that are more efficient to implement in practice (albeit with worse security and privacy guarantees).

The first one (discussed in section~\ref{sec-simple}) operates without additional services in an actively adversarial model, but nevertheless protects user privacy better than plain contact discovery protocols: it prevents adversaries from recovering contacts if little prior knowledge about contacts is known, and in particular prevents the server from recovering the social graph. The second protocol (see section~\ref{sec-keyserver}) significantly improves the privacy protection offered, because it also prevents an active adversarial server or third party to test even a suspicion of two users being a contact. It is actually quite similar to the main protocol presented in section~\ref{sec-protocol}, requiring an online key server instead.

We briefly sketch the protocols and argue why they have the properties we claim they have, without a full formal model or proof.

\subsection{Security and privacy definitions}

The first protocol in this appendix operates in a weaker setting, where guessing a particular member's identifier or a potential contact matters. Therefore, apart from the general threat model discussed above, for the protocols\footnote{%
  We could have proven the second protocol correct in the more strict setting used in the proof of correctness for the certified identifiers based protocol presented in the main body of this paper, but choose not to for easy of presentation. 
}
discussed in this appendix we need to assume that adversaries will use prior knowledge to maximise their chance of success. In particular, it is realistic to assume an adversary knows the list of phone numbers in use, and knows the phone number of several persons of interest. It may also have some knowledge regarding the likely contacts of a person of interest.

When defining the requirements below, we need to consider adversarial behaviour of both members an the server. In the definitions, $X$ denotes such a set of colluding adversarial entities, which for all definitions given below is a subset of the active members and may include the server. We have the following security requirement.
\begin{definition}[Security]
  \label{def-security}
  Let $A$ and $B$ in $V$ be arbitrary users such that $A \in \members$ follows the protocol.
  If $B \notin \members$ or $B \notin \contacts{A}$ then $X$ cannot force
  $B \in \out{A}$.
\end{definition}
Note that this definition is a slight relaxation of the correctness requirement, as it does not take into account whether $A \in \contacts{B}$. The reason being that if this is not the case, $B$ could easily pretend $A$ is one of its contacts. Also note that the \emph{matching} server does not know or control $\members$.

We have also have the following two privacy requirements (where we write ``$X$ does not learn $P$'' as a shorthand for ``entities in $X$ cannot decide whether $P$ holds by observing a protocol run or by engaging in a protocol run, restricted by the threat model and security assumptions discussed below'').

We first define \term{membership privacy}.
\begin{definition}[Membership privacy]
  \label{def-memberprivacy}
  If $X \cap \contacts{A} = \emptyset$, then $X$ does not learn whether $A \in \members$, for any $A \notin X$ of its choosing.  
\end{definition}
Note that the definition of membership privacy only restricts users that do \emph{not} appear in the contact list of a user $A$ to discover whether $A$ is actually a member of the service. In other words, we do not consider it a violation of membership privacy if some $Z \in \contacts{A}$ discovers $A \in \members$ even if $A \notin \contacts{Z}$, as the fact that $A$ included $Z$ as a contact signals that $A$ allows $Z$ to discover them, and $Z$ can always pretend that $A \in \contacts{Z}$. Also note that $A \notin X$ implies $A$ is honest. The definition also applies to $X$ including the server (which is not a member of $\contacts{A}$ by definition).

We next define \term{contact privacy}.
\begin{definition}[Contact privacy]
  \label{def-contactprivacy}
  $X$ does not learn whether $B \in \contacts{A}$ (or not), for any $A \notin X$ and $B \notin X$ of its choosing.
\end{definition}
Note that this definition sidesteps the problem of $B$ itself being able to detect whether $B \in \contacts{A}$ as it can always pretend that $A \in \contacts{B}$ as discussed above. In other words, this is not considered a violation of contact privacy. Again $A$ and $B$ are honest by definition.

\subsubsection{On the choice of definitions}

As is often the case, there are different ways how one might define relevant requirements that capture all the essential aspects of a particular problem.\footnote{%
  Note that this is much less of a problem in the ideal versus real model used in the main body of the paper as the ideal model captures all the intuitive properties you would want the protocol to have 'for free'.
}
In our case we could have defined a weaker version of contact privacy as ``$X$ does not learn an $A, B \notin X$ such that $A \in \contacts{B}$'' (and a similarly weaker version of membership privacy). Except for contact privacy in the first protocol, that turns out not to matter much (as we will see further on).

Although membership privacy and contact privacy appear to be two entirely different notions at first sight, they are in fact related. If an entity $X$ would be able to break contact privacy, it can test whether
$B \in \contacts{A}$, while by definition $X \cap \contacts{A} = \emptyset$. But if it discovers $B \in \contacts{A}$ then surely $A \in \members$: if $A$ is not participating in a run of the mutual contact discovery protocol, it cannot leak information about its contacts. Of course, to test such a suspicion, $X$ needs to guess the right $B$ (if it guesses wrong, \ie a $B \notin \contacts{A}$, then the negative answer of the contact privacy `oracle' is useless as an indication of whether $A \in \members$). The success of the adversary therefore depends on its knowledge of possible contacts, and we see that membership privacy weakly implies contact privacy.

The other way around, suppose an entity $X$ would be able to break membership privacy, and determine whether $A  \in \members$ while $X \cap \contacts{A} = \emptyset$. No general statement about contact privacy can be made in this case. For example, a system can maintain contact privacy while every member $A \in \members$ broadcasts this fact to the world. On the other hand, this does appear to be a pathological case: the idea of mutual contact discovery is that the fact that a user is a member is hidden for anybody except their contacts. As a result one would expect that the only way to break membership privacy is by learning something about the contacts of $A$.

We conclude that even though membership privacy and contact privacy are related, they are sufficiently different to warrant separate study.

\subsection{Notation}

We use a concatenation operator $\concat$ where we assume that the resulting binary string $x \concat y$ can be unambiguously decomposed into its parts $x$ and $y$. In other words, there do not exist different $x'$ and $y'$ such that $x \concat y = x' \concat y'$.

The protocols below uses a \term{key derivation function} as a primitive.
A key derivation function $\KDF{\cdot}: \{0,1\}^{*} \mapsto \{0,1\}^\secpar$~\autocite{RFC2898} (where $\secpar$ is the security parameter) is a cryptographic hash function that is not only hard to invert (and is pre-image and collision resistant) but also takes a `significant' time to compute. This means the function cannot be inverted in practive over relatively small, low entropy, domains (because computing a dictionary is prohibitively expensive). A good choice would be scrypt~\autocite{RFC7914} or Argon2~\autocite{RFC9106} which won the password hashing competition in 2015.\footnote{%
  See \url{https://www.password-hashing.net}, accessed 2021-10-15.
}
Indeed the problem we face in this paper is very similar to that of protecting passwords stored in password files, as user chosen passwords typically have a low entropy as well.

Usually, key derivations functions (KDFs) include a parameter to tune their time complexity (for example by setting the number of iterations). This can be tweaked to achieve a decent security level for a particular input domain.

\section{A simple protocol}
\label{sec-simple}

Define a token between users $A,B \in \sgu$ as
\[
\mutuall{A}{B} = \KDF{A \concat B}~.
\]
Then the following is a protocol for mutual contact discovery between a member $M \in \members$ and the matching server $\matchingserver$.
\begin{itemize}
\item
  In the \emph{submission phase} each member $M$, for each $A \in \contacts{M}$, computes $\mutuall{A}{M}$. $M$ sends this value to the matching server, which stores it in $S$.
\item
  In the \emph{query phase} each member $M$ now sends $\mutuall{M}{A}$ (\ie with $A$ and $M$ reversed) for all $A \in \contacts{M}$ to the server. The server returns whether $\mutuall{M}{A} \in S$. If so, $M$ records that $A$ is discovered as a mutual contact, adding $A$ to $\out{M}$.
\end{itemize}

\subsection{Informal analysis}

The input to the KDF is the concatenation of two identifiers (typically phone numbers). For a messaging service with a national reach the set of phone numbers is typically several million in size. For a global system like WhatsApp or Signal, this is more like several billion. As mentioned before, contact lists can contain several hundreds of entries.

Let us consider the smaller, national, setting (which is therefore is the most risky). Let's say there are $2^{24} \approx 16,7$ million possible phone numbers. Then (because of concatenation) the input domain to the KDF contains at least $2^{48}$ possible elements (we ignore the round number here). Let us assume that the average customer hardware is a million ($2^{20}$) times slower than the fastest hardware available to the attackers. And let us assume that a very connected member has a contact list of at most several hundred entries (say $2^8$), which each need to be processed by the KDF. Then the effort for an adversary to create a complete dictionary,\footnote{%
  Such a dictionary would actually be huge to begin with: it would be $2^{48}$ bits, which equals roughly $32$ TB.
}
is $2^{48}/(2^{20} \times 2^8) = 2^{20}$ times higher than the effort for a member to participate in the protocol.

Using a KDF thus (in practice) limits parties without any prior knowledge to try all possible inputs and discover relevant ones. Parties with (specific enough) prior knowledge are less constrained and may thus break the privacy properties of the scheme. In the informal analysis presented below we say protection is `limited' in this case.

\paragraph{Correctness}

First of all observe that when $A \in \contacts{B}$ and $B \in \contacts{A}$, \ie when $A$ and $B$ \emph{are} mutual contacts, and when both engage in the protocol during an even and following odd slot, both will be notified of being a mutual contact. In the submission phase $A$ sends $\mutuall{B}{A}$ to the server, who stores it in $S$. In the query phase that follows $B$ sends $\mutuall{B}{A}$ to the server, who detects that this value already is in $S$. 

\paragraph{Security}

To what extent does the protocol satisfy definition~\ref{def-security}?
Note that $A$ only queries the server in the query phase for $B \in \contacts{A}$, so security is maintained if that condition doesn't hold. However, if $B \in \contacts{A}$ while $B \notin \members$, then an adversarial server $X$ simply can reply yes and pretend there is a match to any of $A$'s queries. An adversarial member $X$ can compute and send $\mutuall{A}{B}$ in the first phase (even when $A \not\in \contacts{B}$). In the second phase, given that $B \in \contacts{A}$, $A$ will send $\mutuall{A}{B}$ to the server. The server replies that there is a match, and hence $B \in \out{A}$. Security is therefore only guaranteed when $B \notin \contacts{A}$.

\paragraph{Membership privacy}

For membership privacy (see definition~\ref{def-memberprivacy}) we need to prove that any entity $X \notin \contacts{A}$ does not learn whether $A \in \members$, for any $A \neq X$ of its choosing. In fact, membership privacy only holds in a limited sense, when the adversarial entity $X$ is a member or the server. $X$ needs to guess a $Y$ such that $Y \in \contacts{A}$, and then compute $\mutuall{Y}{A}$. If $X$ is an ordinary member it needs to send this value to the server in the query phase, hoping that the server returns true because $\mutuall{Y}{A} \in S$ as indeed $Y \in \contacts{A}$. If $X$ is the server it can simply test this directly. By the properties of the KDF used to compute $\mutuall{Y}{A}$, $X$ has only a limited number of tries, making it infeasible in general. 

\paragraph{Contact privacy}

The protocol does not satisfy contact privacy (definition~\ref{def-contactprivacy}): any member (or the server) can construct
$\mutuall{A}{B}$ and either inspect $S$ (the server) or submit it in the query phase (any other member) to test whether $A \in \contacts{B}$.

It would satisfy contact privacy in a weaker sense where the adversary would be required to find $A$ and $B$ such that $A \in \contacts{B}$. As this involves testing several guesses, the success of the adversary would be limited depending on its knowledge of possible contacts and the amount of resources it is willing to spend.

\subsection{Final comments}

Observe that the server learns the total number of matches. Because a fresh anonymous circuit is used to submit or query a value, the server does \emph{not} learn how many matches a particular member has.

Also observe the use of a key derivation function: it prevents the adversary from trying out all possible combinations of inputs (which would have been feasible when using an ordinary hash function instead), especially because the input to the key derivation function in this protocol is the concatenation of two identifiers, whereas traditionally only a single identifier is hashed. This increases the uncertainty for the adversary considerably.

All in all we stress that even given the obvious shortcomings of this protocol, it is more privacy friendly than the hash-based approach for normal, one-sided contact discovery, for several reasons.
\begin{enumerate}
\item
  To be discovered by another member $A$, $B$ needs to have $A$ as a contact, or $A$ needs to guess a contact of $B$.
\item
  It therefore prevents adversaries from discovering members if little prior knowledge about contacts is known.
\item
  In particular it prevents the server from learning the identifiers of non-members or recovering the social graph from the information that it receives.\footnote{%
    In all fairness it should be mentioned that even for one-sided, traditional, contact discovery the process can be made such that the identity of the entity starting the contact discovery process remains anonymous, therefore also preventing the reconstruction of the social graph by the server.
  }
\end{enumerate}
On the other hand, this particular mutual contact discovery does have a significant privacy leak: any member can test whether $B \in \contacts{A}$. This is typically not possible (by definition) in one-sided contact discovery protocols, as information about $\contacts{A}$ has no bearing on the information exchanged between any other member and the server. In other words there is an inherent trade-off between one-sided contact discovery and mutual contact discovery: the first protects contact privacy better, while the latter protects membership privacy better. 

Finally note that many of the weaknesses in this protocol are caused by the fact that any member can compute and submit/query $\mutuall{A}{B}$ for arbitrary $A$ and $B$. The use of certified identifiers by the protocol in the main body of this paper makes this impossible. An alternative, more efficient, approach that prevents an arbitrary member to compute such a value is used in the next protocol section.

\section{Protocol using a key server}
\label{sec-keyserver}

The crucial observation is that in order to prevent the server to verify a guess, the values used to discover a connection need to involve a secret that the server does not know. As it turns out, that same secret can be used to ensure that users cannot verify a guess for a connection between two arbitrary users. The same insight underlies the main certified identities protocol presented earlier. The following protocol is inspired by, but different from, older match making protocols~\autocite{baldwin1985matchmaking,meadows1986matchmaking}.

Assume there is a key server, independent from the matching server.
This is not an unrealistic assumption, as most messaging services run such a key server anyway to allow users to exchange messages end-to-end encrypted. Alternatively, one could leverage the existing infrastructure of mobile phones (or any other existing infrastructure that provides the identifiers used to discover contacts) and let the mobile network operator maintain such a key database.

Let $\pubkey{A} = g^{\privkey{A}}$ for some generator $g$ over a group in which the Computational Diffie-Hellman (CDH) problem is hard~\autocite{DifH76}.
Let user $A$ generate such a key pair $\privkey{A},\pubkey{A}$ when joining the system, and submit its public key $\pubkey{A}$ to this key server. We assume the key server can verify the identity of users submitting keys (to prevent the adversary from submitting keys for users). When asked for a key of a user that does not exist, the key server generates a random public key and stores it for that user.\footnote{%
  We assume the key server does not collude with the matching server, or else the matching server may be able to detect contacts of unenrolled users as it controls the private key.
}
Subsequent requests for the key of that same (unenrolled) user return the same key. All requests for keys should take the same time to serve. This way, users cannot use the key server to test whether users are enrolled or not.

Consider the protocol using certificates from section~\ref{sec-protocol}. Instead of certificates we use public keys to compute the necessary tokens. That is, we define
\[
\token{A}{B} = \pubkey{A}^{\privkey{B}} = g^{\privkey{A}\privkey{B}}~.
\]
and run the protocol as before, with the following modification:
when each member $M$, for each $A \in \contacts{M}$, computes $\token{A}{M}$ it first retrieves $\pubkey{A}$ from the key server.

\subsection{Performance}

Note how the server is not involved in any complicated computations: it simply stores and looks up values, and answers queries. From the server side this is as good as it could possibly get.

In each round, a user $B$ needs to create a key pair (as explained this involves one exponentiation in a prime-order subgroup of $\Z^*_p$, where $p$ itself is prime) and for each $A \in \contacts{B}$ retrieve $\pubkey{A}$  from the key server and compute $\token{A}{B} = \pubkey{A}^{\privkey{B}}$ (again one exponentiation in a prime-order subgroup of $\Z^*_p$). This is some work, but not excessive: the private set intersection protocols discussed in section
\ref{sec-sota} each require the user to do at least the same amount of work.\footnote{%
  For example, the OPRF construction used in the efficient protocol of Chen~\etal~\autocite{chen2018psi} also uses an exponentiation for each user input.
}

\subsection{Informal analysis}

We again argue the protocol satisfies all required properties in an informal fashion here.

As all network communication is assumed to be anonymous, the key server does not learn the identity of the user requesting a particular public key. As a result the key server cannot reconstruct the social graph. It does learn the number of people interested in contacting a particular someone, though. It is for this reason that we settled for the certified identities based protocol as the main protocol because the issuer is offline and therefore limits the risk of tracing users.

In what follows, we again only consider a single adversarial user or a single adversarial server. As will become clear during the analysis, collusion among several adversarial entities does not help them.

Note that no Diffie-Hellman instance (\ie $\token{A}{B}$ in the protocol) is ever sent in the clear. This in turn implies that the Decisional Diffie-Hellman problem is irrelevant here and that we only need to rely on the hardness of the Computational Diffie-Hellman (CDH) problem.

\paragraph{Correctness}

Correctness immediately follows from the fact that $\token{A}{B} = \token{B}{A}$ and the analysis of the original protocol in section~\ref{sec-protocol}.

\paragraph{Security}

The protocol satisfies definition~\ref{def-security}. For suppose $B \in \contacts{A}$ but $B \notin \members$. Then adversarial $X$ (whether it is the server or a user different from $B$) cannot forge the value $\mutualworkB{A}{B}$ unless it knows $\token{A}{B}$, which is never revealed. Because of the CDH assumption it also cannot compute it.\footnote{%
    Note that this only holds if we assume that the matching server and the key server do not collude; if not, then for unenrolled user $B$ the servers actually know the private key and \emph{can} compute $\token{A}{B}$, and as a result convince $B \in \out{A}$ for those $A$ that have $B \in \contacts{A}$. So if the key server and the matching server collude, protection is limited. The same caveat applies to membership and contact privacy.
  }

\paragraph{Membership privacy}

The protocol satisfies membership privacy (see definition~\ref{def-memberprivacy}). Consider an adversarial $X \notin \contacts{A}$, for any $A \neq X$ chosen by $X$. To test that $A \in \members$, $X$ needs to guess a non corrupted $B$ such that $B \in \contacts{A}$ and then test this choice. It can only do so by sending $\mutualwork{A}{B}$ to the server in the query phase. However, computing $\mutualwork{A}{B}$ requires knowledge of eiter $A$'s or $B$'s secret, so this is impossible. We conclude that $X$ does not learn whether $A \in \members$.

\paragraph{Contact privacy}

The protocol satisfies contact privacy (definition~\ref{def-contactprivacy}).
To test whether $A \in \contacts{B}$, the adversary needs to compute $\mutualwork{A}{B}$ and either query the server, or (when the server itself is adversarial) test membership in $S$. However, computing $\token{A}{B}$ requires knowledge of either $A$'s or $B$'s secret, so this is impossible.

\subsection{Final comments}

As in the first protocol the matching server learns the total number of matches.
(When a fresh anonymous circuit is used to submit or query a value, the matching server does not even learn how many matches a particular user has.)

The key server learns the number of members and their identity. As discussed before, we need to assume queries to the key server are anonymised (shielding the network address of the requestor) to prevent the key server from also reconstructing the full social graph. The key server does learn the number of users that have a particular user in their contact list, \ie it learns how popular (or unpopular) a user is, by counting how many times a particular key is requested. To prevent the key server from learning this, an (inefficient) approach would be to use private information retrieval~\autocite{chor1998pir}. In any case, note that this is about all the key server learns.

Observe that if the main protocol using certified identifiers would rely on a separate certificate issuer, this issuer also learns the identities of all participating members.

\section{Discussion}

Compared to the protocol used by Signal (that essentially matches hashes of phone numbers found in contact lists), the protocols discussed in the appendix should perform with similar performance. In both our protocols the matching server only performs simple store and lookup operations. User side we expect only a constant factor slowdown (depending on the time required to compute the key derivation function) as both our protocols also only hash (using the key derivation function) entries in the contact lists, and the second protocol requires only a single exponentiation for each entry in the contact list.

We wish to stress that the definition of membership privacy only restricts members that do \emph{not} appear in the contact list of a member $B$ to discover whether $B$ is actually a member of the service. Even in the second protocol it is still possible that a member $A$ guesses a $B$ with $A \in \contacts{B}$ and then test this guess by sending $\tokenx{A}{B}{A}{B}$ to the server in the query phase. $A$'s luck depends on its prior knowledge of such $B$. Note though that $A$ has only a limited number of tries, making it infeasible in general. We do not consider this an attack as the fact that $B$ included $A$ as a contact signals that $B$ allows $A$ to discover them. (Modulo the discussion in section~\ref{sec-intro}.)

Regarding contact privacy for the first protocol note that the definition of contact privacy is strong. A weaker version of contact privacy that requires the adversary to find a pair $A,B$ such that $B \in \contacts{A}$ is actually satisfied by the first protocol in a limited sense (depending on its knowledge and the amount of times it is willing to evaluate the key derivation function).

The first protocol prevents adversaries from recovering contacts if little prior knowledge about contacts is known, and in particular limits the ability of the server from recovering part of the social graph from the information that it receives. The second protocol significantly improves the privacy protection offered, because it also prevents an active adversarial server or third party to test even a suspicion of two members being a contact. To do so, the use of a key server is required. Note that a distinct advantage of the first protocol discussed in this appendix is that it does not rely on such an infrastructural assumption.